\documentclass{elsart}
\usepackage{graphicx}

\newcommand{\lepton}{\ifmmode {l} \else $l$\fi}
\newcommand{\wboson}{\ifmmode {{\mathrm W}^{\pm}} \else
${\mathrm W}^{\pm}$\fi}
\newcommand{\wpair}{\ifmmode {{\mathrm W}^{+}{\mathrm W}^{-} } \else
${\mathrm W}^{+}{\mathrm W}^{-}$\fi}
\newcommand{\zboson}{\ifmmode {{\mathrm Z}^{0}} \else
${\mathrm Z}^{0}$\fi}

\usepackage{amssymb}
\usepackage{amsmath}
\usepackage{mathrsfs}
\usepackage{epsfig}

\begin{document}

\begin{frontmatter}


\title{The sensitivity of cosmic ray air shower experiments for leptoquark detection}

\author[lip]{M.C. Esp\'{\i}rito Santo}
\author[lip,cat]{A. Onofre}
\author[lip]{M. Paulos}
\author[lip,ist]{M. Pimenta \corauthref{cor}}
\author[ist]{J. C. Rom\~ ao}
\author[lip]{B. Tom\'e\thanksref{grant}}
\address[lip]{LIP, Av. Elias Garcia, 14--1, 1000-149 Lisboa Portugal}
\address[ist]{IST, Av. Rovisco Pais, 1049-001 Lisboa, Portugal}
\address[cat]{Universidade Cat\'olica, Figueira da Foz, Portugal}
\corauth[cor]{pimenta@lip.pt, LIP, Av. Elias Garcia, 14--1, 1000-149 Lisboa Portugal.}
\thanks[grant]{FCT grant SFRH/BPD/11547/2002.}

\begin{abstract}
Leptoquarks arise naturally in models attempting the unification of
the quark and lepton sectors of the standard model of particle
physics.  Such particles could be produced in the interaction of high
energy quasi-horizontal cosmic neutrinos with the atmosphere, via
their direct coupling to a quark and a neutrino.  The hadronic decay
products of the leptoquark, and possibly its leptonic decay products
would originate an extensive air shower, observable in large cosmic
ray experiments.  In this letter, the sensitivity of present and
planned very high energy cosmic ray experiments to the production of
leptoquarks of different types is estimated and discussed.
\end{abstract}

\end{frontmatter}

\section{Introduction}
\label{sec:introd}

In this paper, the possibility of leptoquark searches in current
(AGASA~\cite{agasa}, Fly's Eye~\cite{fly}) and future
(Auger~\cite{auger}, EUSO~\cite{euso}, OWL~\cite{owl}) very high
energy cosmic ray experiments is discussed. The approach outlined
in~\cite{excit-our} is closely followed.

Leptoquarks arise
naturally in several models attempting the unification of the quark
and lepton sectors of the Standard Model (SM) of particle physics.
Different leptoquark types are expected, according to their quantum
numbers, which give rise to different coupling strengths and decay
modes, and thus to different cross-sections and final states.  The
fact that leptoquarks provide a direct coupling between a quark and a
lepton, charged or neutral, makes them unique particles, which should
lead to signatures that have been thoroughly searched for at man-made
accelerators.  In fact, in the past years many searches were performed
at colliders around the world~\cite{accel}. Whereas in $ep$ collisions
at HERA leptoquarks could be $s$-channel produced, in all cases
($ep$, $e^+e^-$ and hadron colliders) 
they could arise as $t$-channel mediators of SM-like
processes. If light enough, leptoquarks could arise at accelerators as
final state particles of specific processes. So far, no evidence for
leptoquarks was found, and stringent limits were set at the
electroweak scale. In fact, couplings and masses of leptoquarks are
constrained indirectly by low energy experiment and by the precise
measurement of the $Z^0$ width, and direct and indirect searches 
at accelerators have set constraints at higher energies.
It should be noted that most limits obtained at accelerators are
valid for first family leptoquarks only, as the initial beams 
involve first family charged leptons. This is not the case for
some of the Tevatron results, and will be clearly stated in the
discussion of the results.

Large cosmic ray experiments, covering huge detection areas, are able
to explore the high energy tail of the cosmic ray spectrum, reaching
centre-of-mass energies orders of magnitude above those of man made
accelerators.  Although having poorer detection capabilities and large
uncertainties on the beam composition and fluxes, cosmic ray
experiments present a unique opportunity to look for new physics at
scales far beyond the TeV.  Energetic cosmic particles interact with
the atmosphere of Earth originating Extensive Air Showers (EAS)
containing billions of particles.
 
Energetic cosmic neutrinos, although not yet observed and with very
large uncertainties on the expected fluxes, are predicted on rather
solid grounds~\cite{cosmneut}.  Nearly horizontal neutrinos, seeing a
large target volume and with negligible background from ``ordinary''
cosmic rays, are an ideal beam to explore possible rare
processes~\cite{exoticnu}.  In particular, if the available energies
are high enough, the interaction of cosmic neutrinos with the
atmospheric nuclei should create the ideal conditions for the
production of leptoquarks, with dominance of $s$-channel resonant
production.  As the initial beam must contain all three neutrino flavours,
one expects the production of leptoquarks of first, second and
third family.
The produced leptoquarks are expected to decay promptly
into a quark and a charged or neutral lepton. The branching ratio into
the charged and neutral decay mode depends on the leptoquark type.


\section{Leptoquark production and decay}
\label{sec:lq}

Leptoquarks are coloured spin 0 or spin 1 particles with non-zero
baryon and lepton quantum numbers. They are predicted by different
extensions of the SM.  In this paper, we follow the conventions and
theoretical framework formulated in \cite{buchmuller:1986zs}, where the
most general $SU(3) \times SU(2) \times U(1)$ invariant Lagrangian is
given for each family as,
\begin{eqnarray}
  \label{eq:1}
  \mathcal{L}&=&\left( g_{1L}\, \bar{q}^c_L i \tau_2 \ell_L 
      +g_{1R}\, \bar{u}^c_R e_R \right) S_1 
+ \tilde{g}_{1R}\, \bar{q}^c_R e_E \tilde{S}_1 
+ g_{3L}\, \bar{q}^c_L i \tau_2 \boldsymbol{\tau} \ell_L \mathbf{S_3}\nonumber \\
&&+ \left(g_{2L}\, \bar{d}^c_R \gamma^{\mu} \ell_L 
+ g_{2R}\, \bar{q}^c_L \gamma^{\mu} e_R \right) V_{2\mu}
+ \tilde{g}_{2L}\, \bar{u}^c_R \gamma^{\mu} \ell_L \tilde{V}_{2\mu}
+ \tilde{h}_{2L}\, \bar{d}_R \ell_L \tilde{R}_2
\nonumber \\
&&+\left(h_{2L}\, \bar{u}_R \ell_L + h_{2R} \bar{q}_L i \tau_2 e_R
      \right) R_2 
+ \left( h_{1L}\, \bar{q}_L \gamma^{\mu} \ell_L + h_{1R}\, \bar{d}_R
      \gamma^{\mu} e_R \right) U_{1\mu}\nonumber \\ 
&&+\tilde{h}_{1R}\, \bar{u}_R \gamma^{\mu} e_R \tilde{U}_{1\mu}
+h_{3L}\, \bar{q}_L \boldsymbol{\tau} \gamma^{\mu} \ell_L
      \mathbf{U_{3\mu}} + \mathbf{h.c.} \ .
\end{eqnarray}
We can see that there are eighteen leptoquarks per family: nine scalars and
nine vectors. We also notice that some of these are grouped into weak
isospin doublets or triplets, which is referred to by its index (1 for
scalar, 2 for doublet, 3 for triplet). As in \cite{buchmuller:1986zs}
we assume that the couplings are diagonal in generation space.

It is assumed that only one of the chiral
coupling constants is non-zero. In the case of $e p$ collisions, it
was shown~\cite{buchmuller:1986zs}, that qualitatively similar results
could be obtained either by taking $\lambda_R=0$ or $\lambda_L=0$,
where $\lambda$ represents the coupling relevant to the leptoquark in
question ($g_i,\tilde g_i,h_i, \tilde h_i$) as defined in
Eq.~(\ref{eq:1}). However for $\nu p$ collisions, only the case
$\lambda_L\not=0$ is of interest otherwise the neutrino will not
couple to the leptoquarks. So in the following we consider
$\lambda_R=0$.
From the above Lagrangian we can easily derive the decay modes and
coupling constants for neutrino induced leptoquark production. This is
shown in table~\ref{table:interactions}, where the listed leptoquarks
can be of first, second or third family (family indices are ommited) 
and $D=d,s,b$ and $U=u,c$,
as we neglect the top parton distribution function (PDF).  Unless
explicitly stated, in all our results we consider the leptoquarks for
which the factor $\sqrt{2}$ in the couplings is not present. For these
other leptoquarks the results can be obtained by a simple rescaling.
In the table, the possible decay modes (charged and neutral 
or neutral only) of each leptoquak type are also indicated.

\begin{table}[ht]
\centering
\begin{tabular}{|c|c|c|c|}\hline
Leptoquark & Interaction & Decay & Coupling\\ \hline
\multicolumn{4}{|c|}{\bf Scalars}\\ \hline 
$S_1$ & $\nu D $ & $\nu D,\ell^-U$ & $-g_{1L},g_{1L}$\\ 
$S_3^3$ & $\nu D$ & $\nu D,\ell^-U$ & $-g_{3L},-g_{3L}$ \\
$S_3^-$ & $\nu U$ & $\nu U$ & $\sqrt 2 g_{3L}$ \\
$R_2^U$ & $\nu \bar U$ & $\nu \bar U$ & $h_{2L}$ \\
$\tilde R_2^U$ &$\nu \bar D$&$ \nu\bar D $&$\tilde h_{2L}$  \\ \hline
\multicolumn{4}{|c|}{\bf Vectors}\\ \hline 
$V_{2\mu}^U $ & $\nu D $ & $\nu D$ & $g_{2L}$\\ 
$\tilde V_{2\mu}^U

$ & $\nu U$& $\nu U$ & $\tilde g_{2L}$ \\
$U_{1\mu}$ & $\nu \bar U $ & $\nu \bar U,\ell^-D$ & $h_{1L},h_{1L}$\\ 
$U_{3\mu}^3$ & $\nu \bar U$ & $\nu \bar U,\ell^-\bar D$ & $ h_{3L},-h_{3L}$ \\
$U_{3\mu}^-$ & $\nu \bar D$ & $\nu \bar D$ & $\sqrt 2 h_{3L}$ \\ \hline
\end{tabular}
\caption{Relevant processes for neutrino induced leptoquark
  production, assuming $\lambda_R=0$. Here $D=d,s,b$, $U=u,c$,
$\ell=e,\mu,\tau$ and $\nu=\nu_e,\nu_\mu,\nu_\tau$.}
\label{table:interactions}
\end{table}

The processes shown in table~\ref{table:interactions} can thus occur for
leptoquarks of the first, second or third family. For our range of
energies we are probing very small values of $x$ and the cross
sections are comparable for the three families. 
In the following, first family leptoquarks will sometimes be taken
as a case study, but all families will be considered in the result
derivation.

\begin{figure}[htbp]
  \centering
  \includegraphics{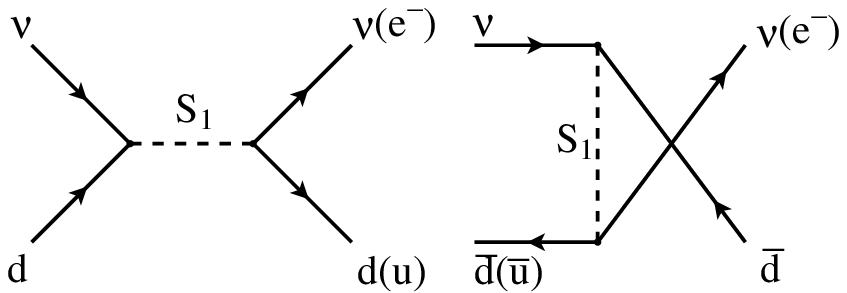}
  \caption{Lowest order Feynman diagrams for leptoquark production in
    neutrino-quark collisions via s-channel (left) and u-channel
    (right) interactions.}
  \label{fig:LQdiags}
\end{figure}

\begin{figure}[hbtp]
\begin{center}
\setlength{\unitlength}{0.0105in}%
\includegraphics[width=0.65\linewidth]{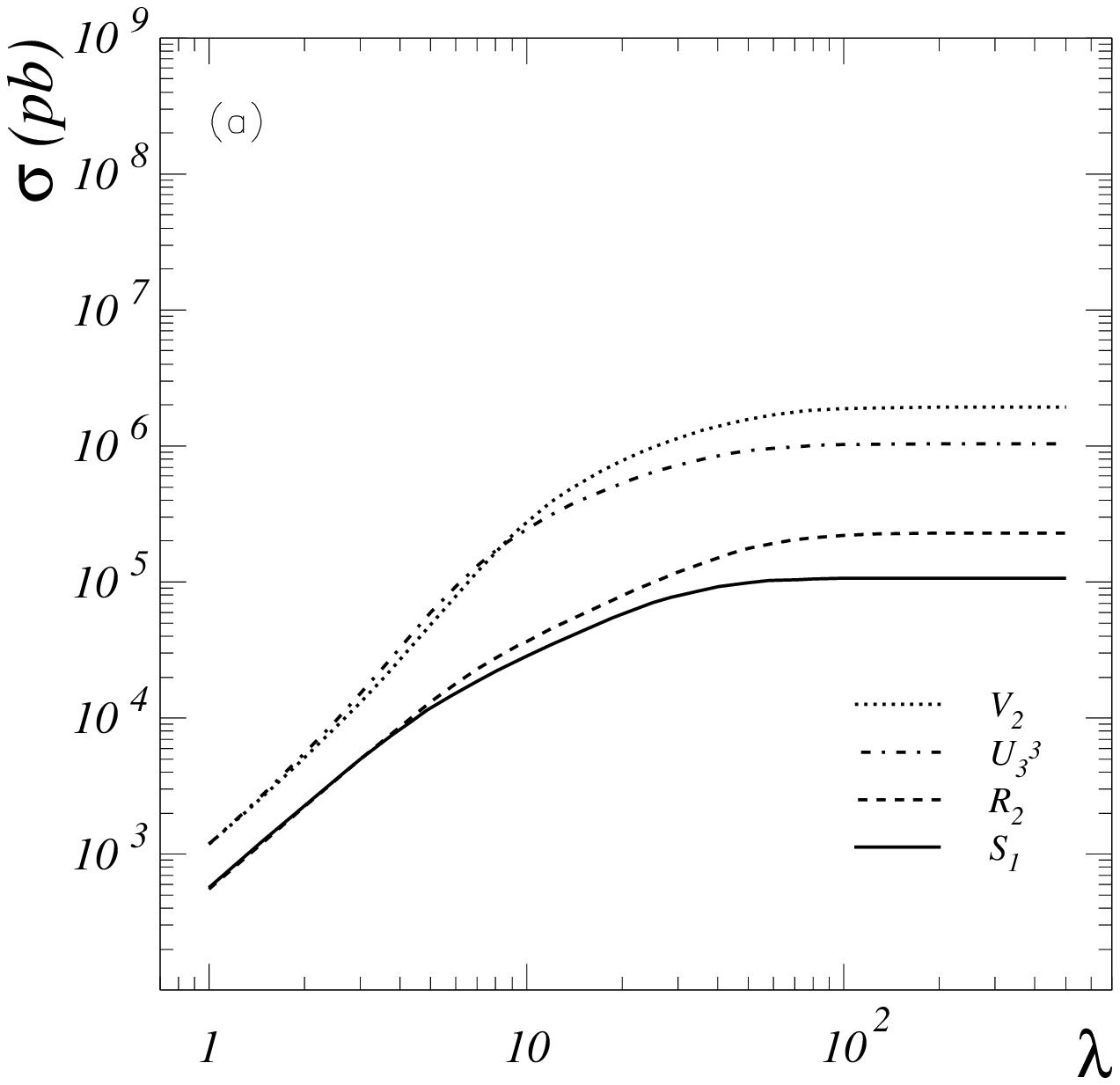}
\includegraphics[width=0.65\linewidth]{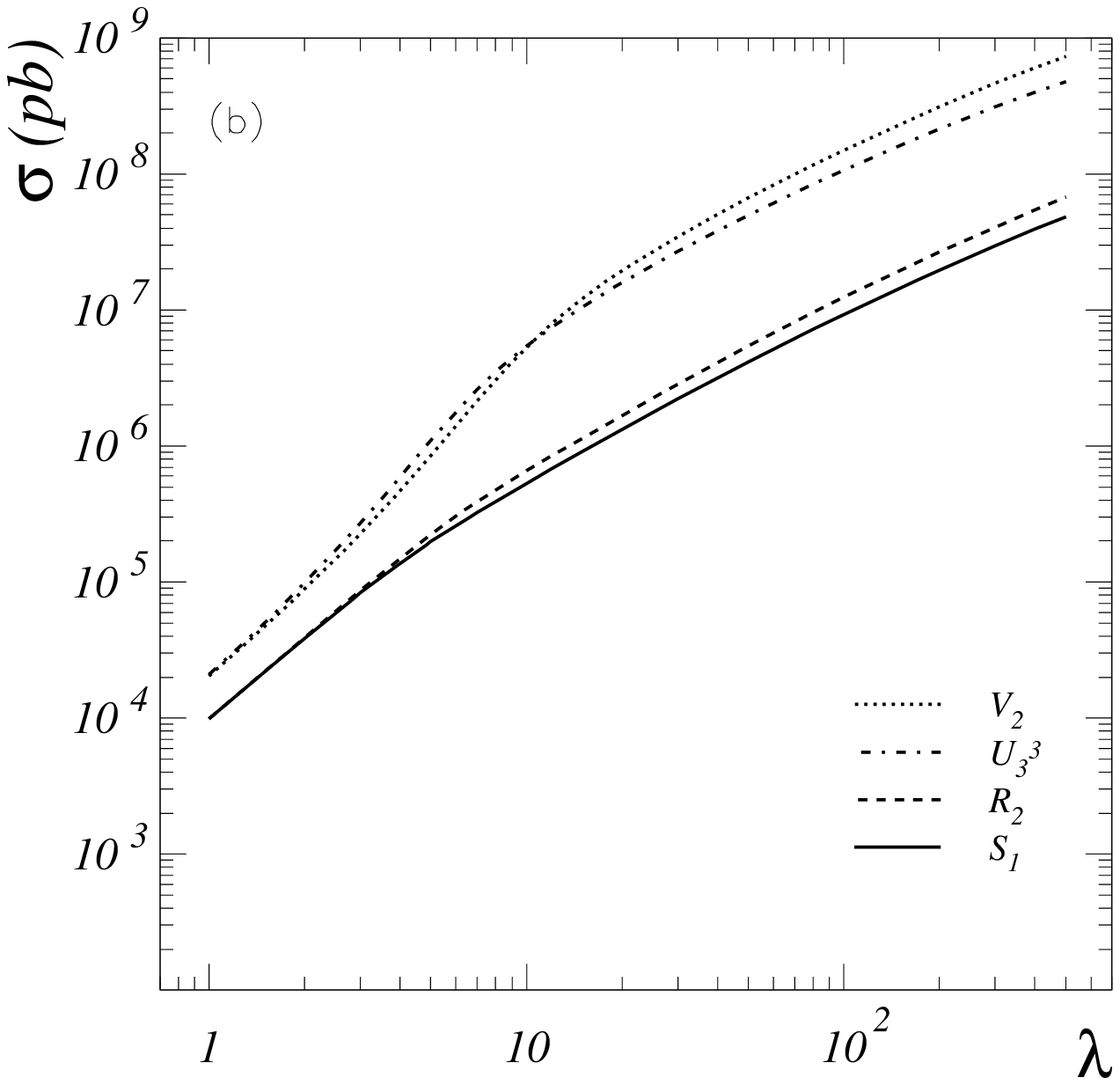}
\caption{Total cross-sections for scalar and vector
  leptoquarks, as a function of the coupling $\lambda$, for
  $M_{LQ}=1$~TeV, and two different neutrino energies: (a)
  $E_{\nu}=10^{17}$~eV, (b) $E_{\nu}=10^{20}$~eV. 
  The upper (lower) lines correspond to the cases of leptoquarks 
  having two (one) decay modes.}
\label{fig:xslambda}
\end{center}
\end{figure}

\begin{figure}[ht]
\begin{center}
\setlength{\unitlength}{0.0105in}%
\includegraphics[width=0.65\linewidth]{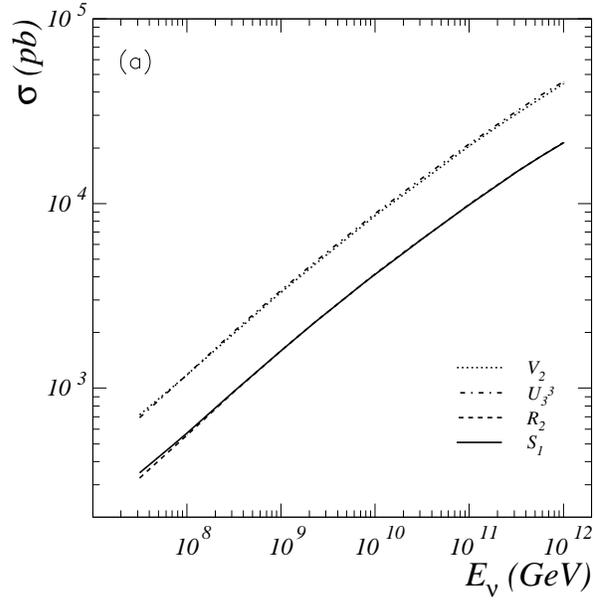}
\includegraphics[width=0.65\linewidth]{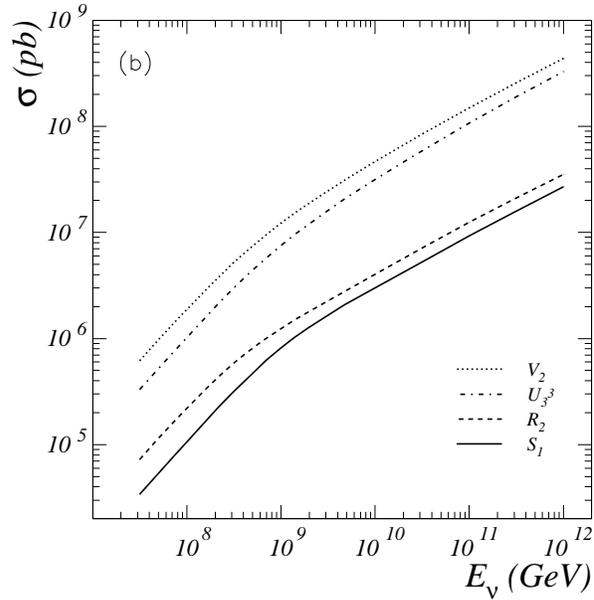}
\caption{Total cross-sections for scalar and vector first family
leptoquarks, as a function of energy for $M_{LQ}=1$~TeV,
and two different values of $\lambda$: (a) $\lambda=1$
and (b) $\lambda=100$.} 
\label{fig:xsEnergy}
\end{center}
\end{figure}

%
All the processes in table~\ref{table:interactions} occur by
$s$-channel. However, flipping the quark to antiquark in each case
leads to an alternative reaction, this time mediated by $u$-channel
leptoquark exchange, as it is shown in Fig.~\ref{fig:LQdiags}.
All the relevant amplitudes are given explicitly
in~\cite{buchmuller:1986zs} for $e q$ interactions and can be easily
adapted to our $\nu q$ case, so we will not repeat them here. The
contribution from $s$-channel resonant production is largely dominant
up to moderate values of $\lambda$.  In fact for values $\lambda \leq
5$ the width is small compared with the leptoquark mass and the
differential cross-section is strongly peaked on the $x$ value
corresponding to the resonance pole, $x=m_{LQ}^2/s$, which gives the
main contribution to the total cross-section. For this range of
$\lambda$ the narrow-width approximation~\cite{buchmuller:1986zs}
\begin{equation}
  \label{eq:3}
  \sigma(\nu p \rightarrow LQ)=\frac{\pi}{4 s}\, \lambda^2 q\left(
  \frac{m^2_{LQ}}{s}\right) \times C_J \qquad C_J=1,2\quad
  \hbox{for}\quad  J=0,1
\end{equation}
explains why the cross-section rises like $\lambda^2$, as it is shown in
Fig.~\ref{fig:xslambda}.
However, for larger values of $\lambda$ the narrow width approximation
can no longer be used and the dependence of the cross-section with
$\lambda$ begins to flatten out, depending on the energy of the
incident neutrino. For these
larger values of $\lambda$ the $u$-channel is no longer negligible and
has to be taken in account. Also, the leptoquarks with one or two
decay modes, indistinguishable in the
narrow with approximation, start to separate for larger values of
$\lambda$. In Fig.~\ref{fig:xsEnergy} we
plot the total cross-section as a function of the energy for different
values of $\lambda$.
As we intend to obtain sensitivities for values of $\lambda$ that can
be above the conditions of applicability of the narrow with
approximation, in this work we always performed the complete
calculations, without any approximation.

\begin{figure}[ht]
\begin{center}
  \begin{tabular}{cc}
\includegraphics[clip,width=0.7\linewidth]{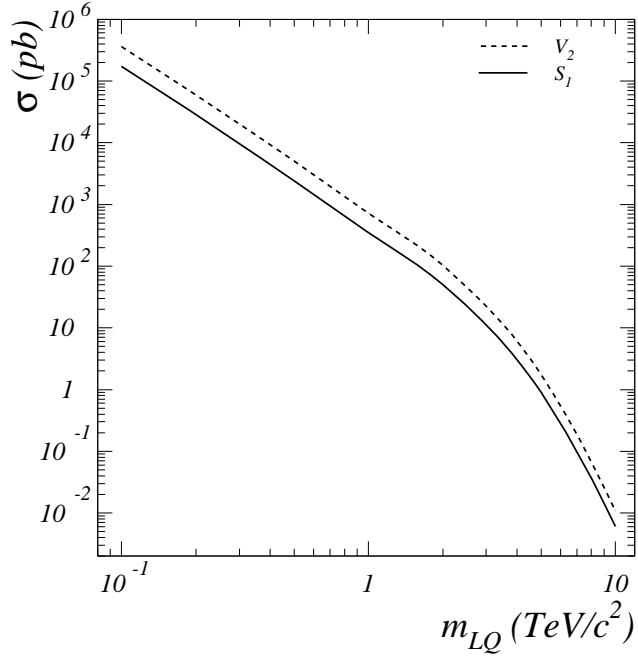}
\end{tabular}
\caption{Total cross-sections for scalar and vector first family
leptoquarks, as a function of the leptoquark mass for $E=10^{20}$~eV
and $\lambda=1$.} 
\label{fig:xsMass}
\end{center}
\end{figure}

The dependence of the total cross-section on the leptoquark mass is 
shown in Fig.~\ref{fig:xsMass} for a scalar and a vector leptoquark.
The curves corresponding to other scalars (or vectors) of the 
same family are identical to these two.

\begin{figure}[htb]
  \centering
  \includegraphics[clip,width=0.7\linewidth]{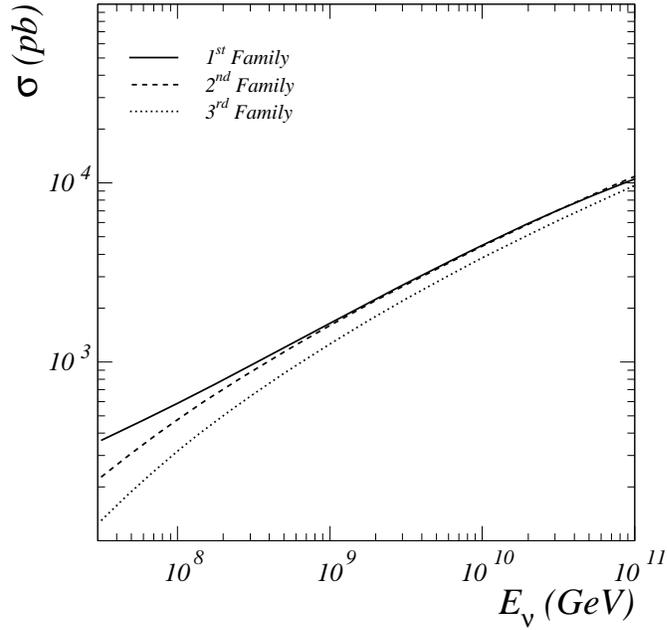}
  \caption{Total cross sections for $S_1$ leptoquarks of different families, 
  as a function of the incindent neutrino energy, for $M_{LQ}=1$~TeV
  and $\lambda=1$.}
  \label{fig:FamilyComparison}
\end{figure}

All the figures above correspond to first family leptoquarks.
In Fig.~\ref{fig:FamilyComparison} the comparison of 
the total cross-section for the
leptoquark $S_1$ of different families is shown as a function of 
incident neutrino energy. As expected, the observed differences 
are attenuated when the energy increases, as mass effects become
less relevant. 

\section{Limits and sensitivities}
\label{sec:sensitivities}

\subsection{Acceptances}

The expected number of observed leptoquark events is given by:
\begin{equation}
\label{eq:nevents}
{\mathscr N} = N_A \int~\frac{d\phi_{\nu}}{dE_{\nu}}~\sigma_{{\nu N}}~{\mathscr A}~\Delta T~dE_{\nu},
\end{equation}
where $d\phi_{\nu}/dE_{\nu}$ is the incident neutrino flux,
$\sigma_{{\nu N}}$ is the appropriate production cross-section,
depending on which leptoquark type is considered, ${\mathscr A}$ is
the acceptance of the experiment for the extensive air showers
produced by these final states, $\Delta T$ is the observation time
interval and $N_A$ is Avogrado's number.  It is assumed that the
attenuation of neutrinos in the atmosphere can be neglected, which is
a safe assumption for total neutrino-nucleon cross-sections in the
range relevant for the present study. For larger values of the
cross-section, the treatment discussed in~\cite{agasa-thesis} should
be applied.  In this work the Waxman-Bahcall (WB)~\cite{wb} bound with
no z evolution, $E_\nu^2 \frac{d\phi}{dE_\nu} = 10^{-8}$ [GeV/cm$^2$ s
sr], is assumed.  The acceptance ${\mathscr A(E)}$ includes both the
geometrical aperture, the target density and the detection efficiency
factors.  The procedure outlined in~\cite{excit-our} was followed to
obtain estimations of the acceptances of the different
experiments~\cite{agasa-a,auger-a,euso-a,owl-a}.  The observation
times were assumed to be: 10 years for Auger, 3 years and 10\% duty
cycle for both EUSO and OWL. For AGASA and Fly's Eye, we followed
reference~\cite{agasa-a}.

As noted in~\cite{excit-our}, the relation between the shower energy
and the primary neutrino energy is process dependent. In the present
case, it depends on the leptoquark type, which determines its decay
branching fractions.  

For the leptoquark charged decay mode (see table~\ref{table:interactions}), 
all the energy of the primary neutrino
will contribute to the development of the extensive air
shower. For the neutral decay mode, on the other hand, only the
hadronic decay products will contribute. 
In general, the actual shower energy
will depend on the branching ratio and on the $d\sigma_{\nu N}/dy$ 
distribution. 
From the
convolution of the two, an average shower energy was obtained.  The
acceptances compiled in \cite{excit-our} were then considered.

\subsection{Sensitivities for $1^{st}$ family leptoquarks}

Using
equation~\ref{eq:nevents}, the sensitivity of the different
experiments to leptoquark production, as a function of the leptoquark
mass, was studied. Requiring the observation of one event, the
sensitivity on the coupling $\lambda$ (see section~\ref{sec:lq}) as a
function of the mass was derived.
The expected sensitivities of the different cosmic ray experiments
for $1^{st}$ family leptoquarks of different types are presented 
in this section.
The assumed observation times are the ones detailed above (and quoted in
the caption of the figure). 

\begin{figure}[hbtp]
\begin{center}
\setlength{\unitlength}{0.0105in}%
\includegraphics[width=0.7\linewidth]{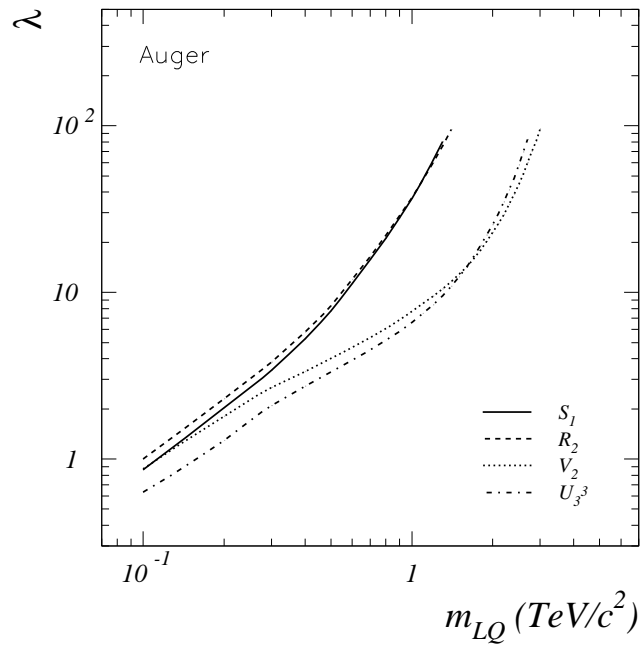}
\end{center}
\caption{Estimated sensitivities of Auger for different
$1{st}$ family leptoquarks, as a function of the leptoquark mass,
for an observation period of 10 years.}
\label{fig:lims-auger}
\end{figure}

Fig.~\ref{fig:lims-auger} shows the sensitivity on the coupling 
$\lambda$ as a function of the leptoquark mass expected in Auger,
for different types of scalar and vector leptoquarks.
As expected, better sensitivities are obtained for vector leptoquarks,
due to the larger cross-sections. Other differences are due to coupling
and branching ratio effects.

\begin{figure}[hbtp]
\begin{center}
\setlength{\unitlength}{0.0105in}%
\includegraphics[width=0.65\linewidth]{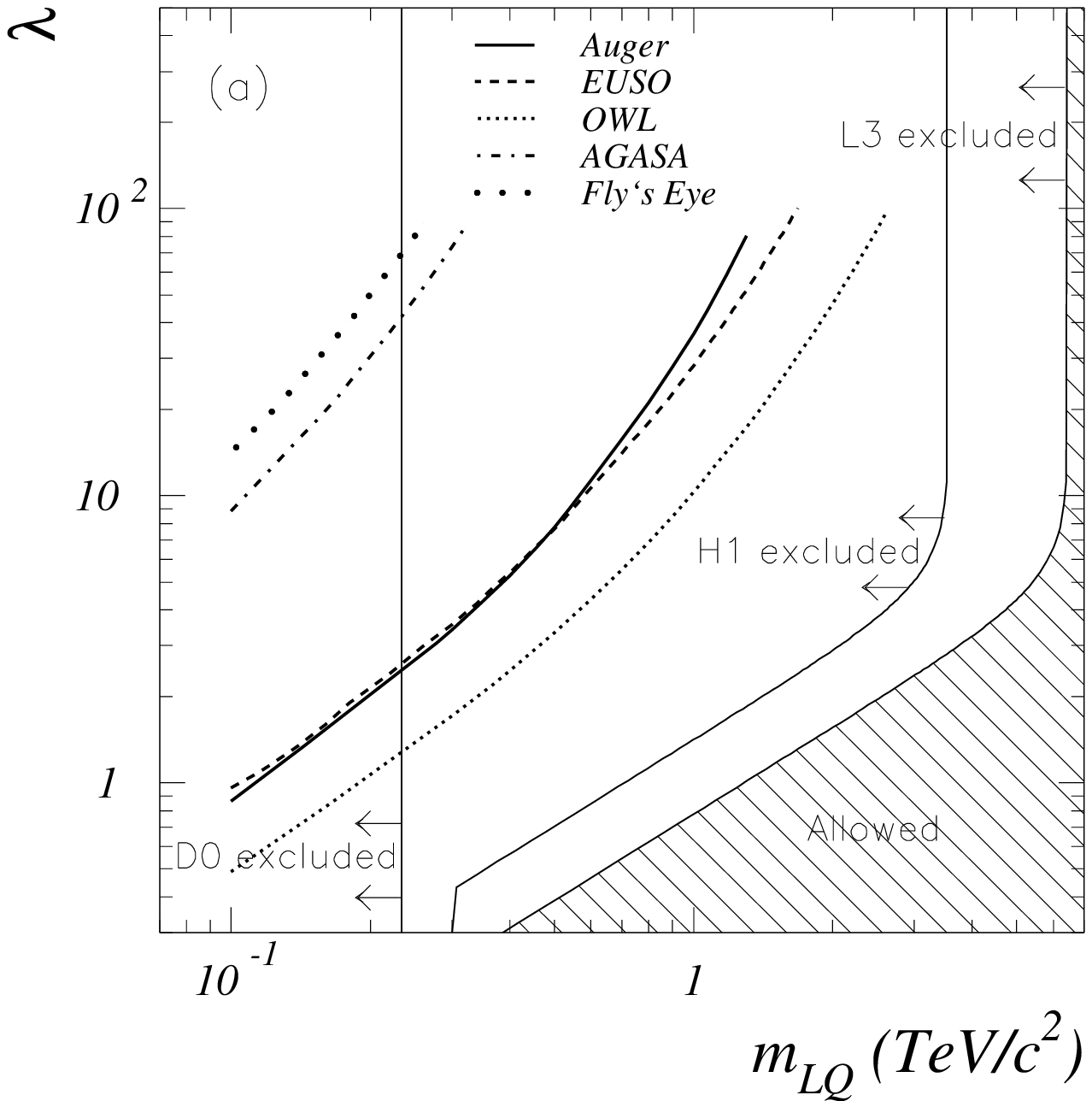}
\includegraphics[width=0.65\linewidth]{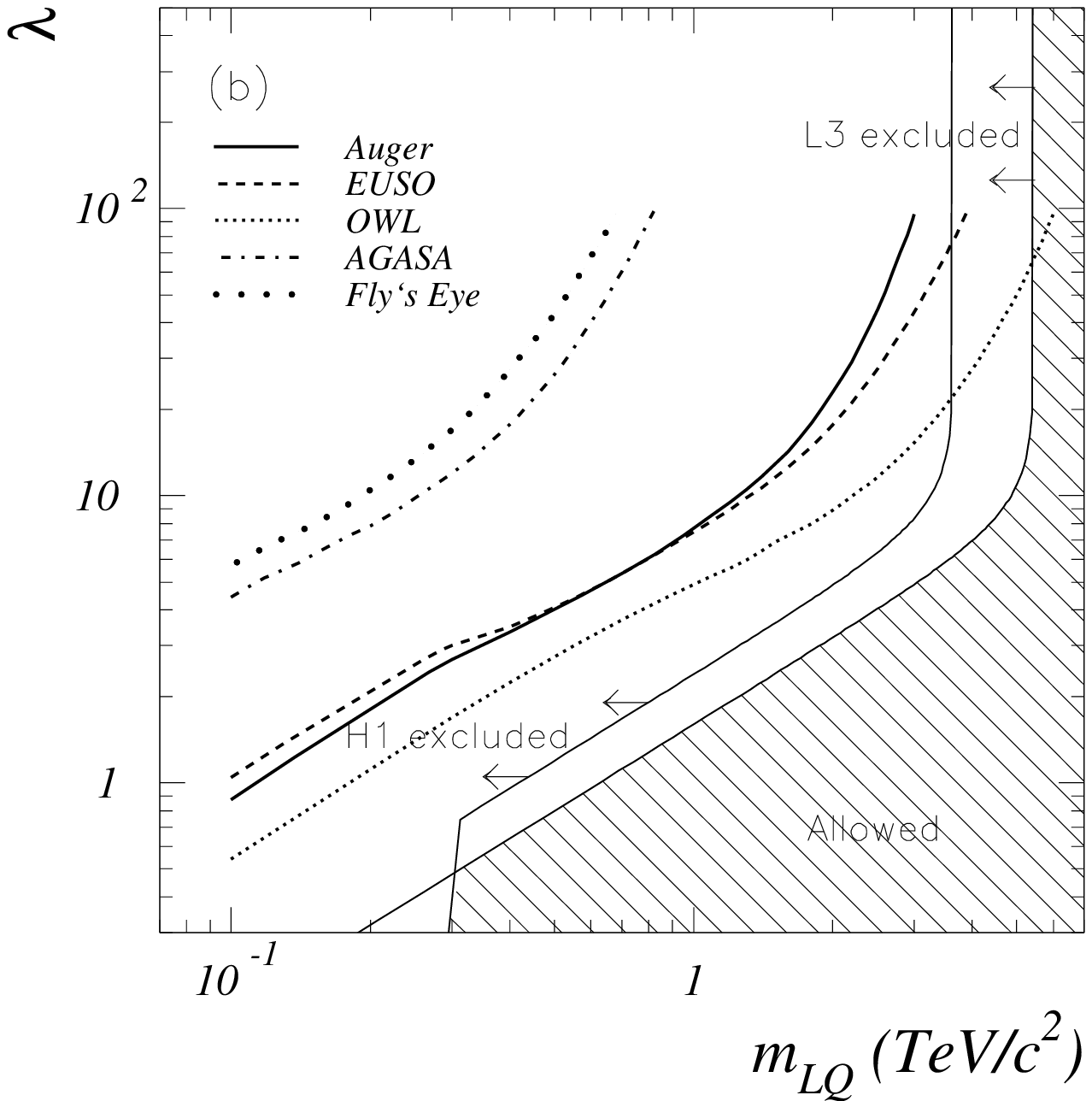}
\end{center}
\caption{Estimated sensitivities of the different cosmic ray
experiments for scalar $S_1$ (a) and vector $V_2$ (b) $1{st}$ family 
leptoquarks, as a function of the leptoquark mass.
The regions excluded by accelerator experiments
are also shown (shaded regions) for comparison.
The observation times were taken as: 10 years for Auger, 
3 years and 10\% duty cycle for both EUSO and OWL. 
For AGASA and Fly's Eye, we followed 
reference~\cite{agasa-a}.
}
\label{fig:lims1}
\end{figure}

Fig.~\ref{fig:lims1} shows the expected sensitivities of the different
cosmic ray experiments, as a function
of the leptoquark mass, for first family scalar and vector leptoquarks 
Limits obtained at accelerators are also shown.  
It can be seen that for first family leptoquarks the powerful
limits obtained at LEP (L3 indirect search) and HERA (which include both 
direct and indirect searches
at H1) exclude the region that could be probed at large cosmic
ray experiments,
for the foreseen acceptances, observation time intervals and fluxes.
As shown, low mass regions 
are also excluded by TEVATRON limits.
Within the same family, the sensitivities for the different leptoquark 
types are within a factor of two.
The indirect limits of H1 and L3 were linearly extrapolated to higher
masses. In the TeV region, however, leptoquark width effects were
taken into account, making these limits weaker.

\subsection{Sensitivities for 2$^{nd}$ and 3$^{rd}$ family leptoquarks}

\begin{figure}[hbtp]
\begin{center}
\setlength{\unitlength}{0.0105in}%
\includegraphics[width=0.65\linewidth]{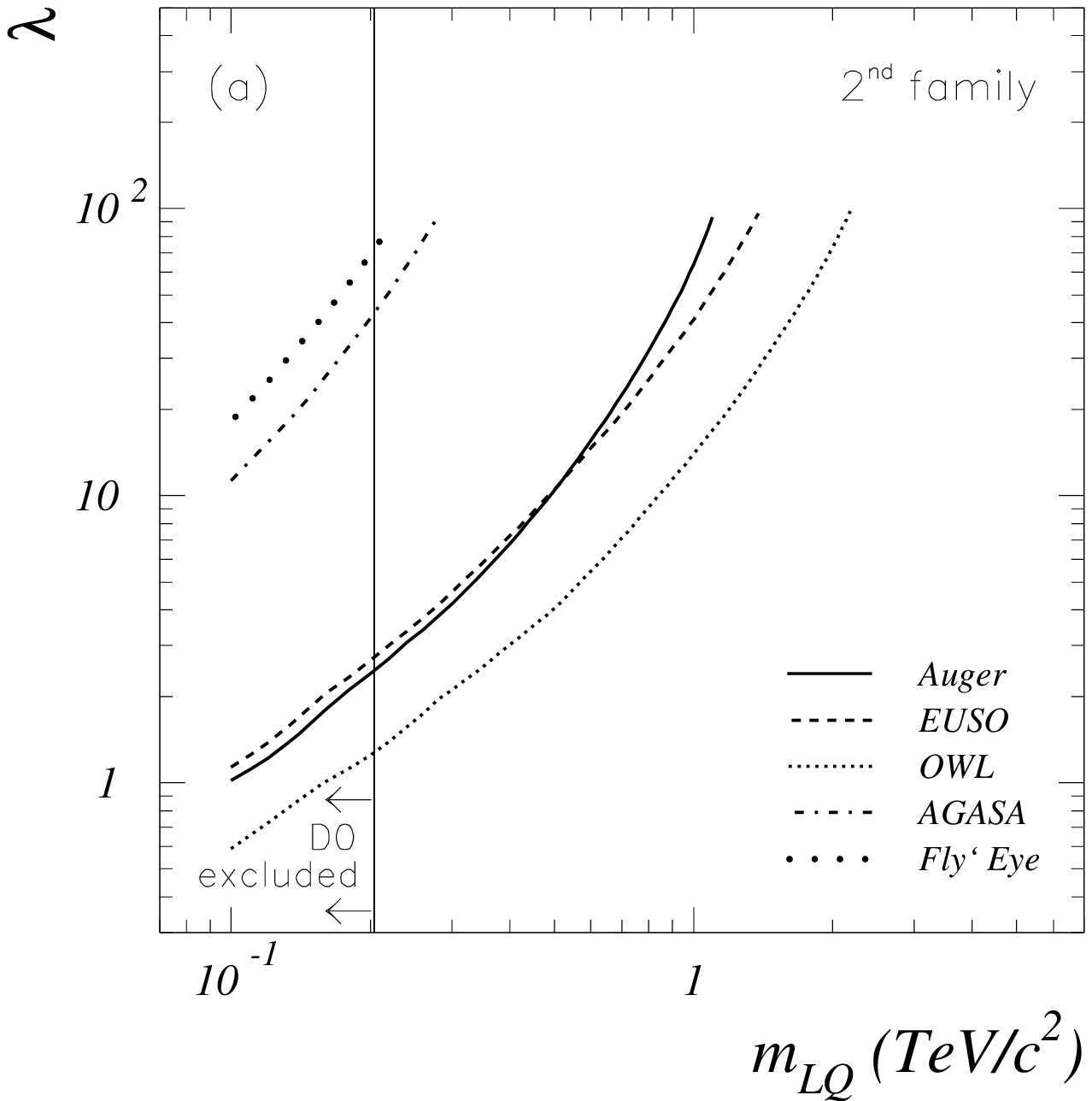}
\includegraphics[width=0.65\linewidth]{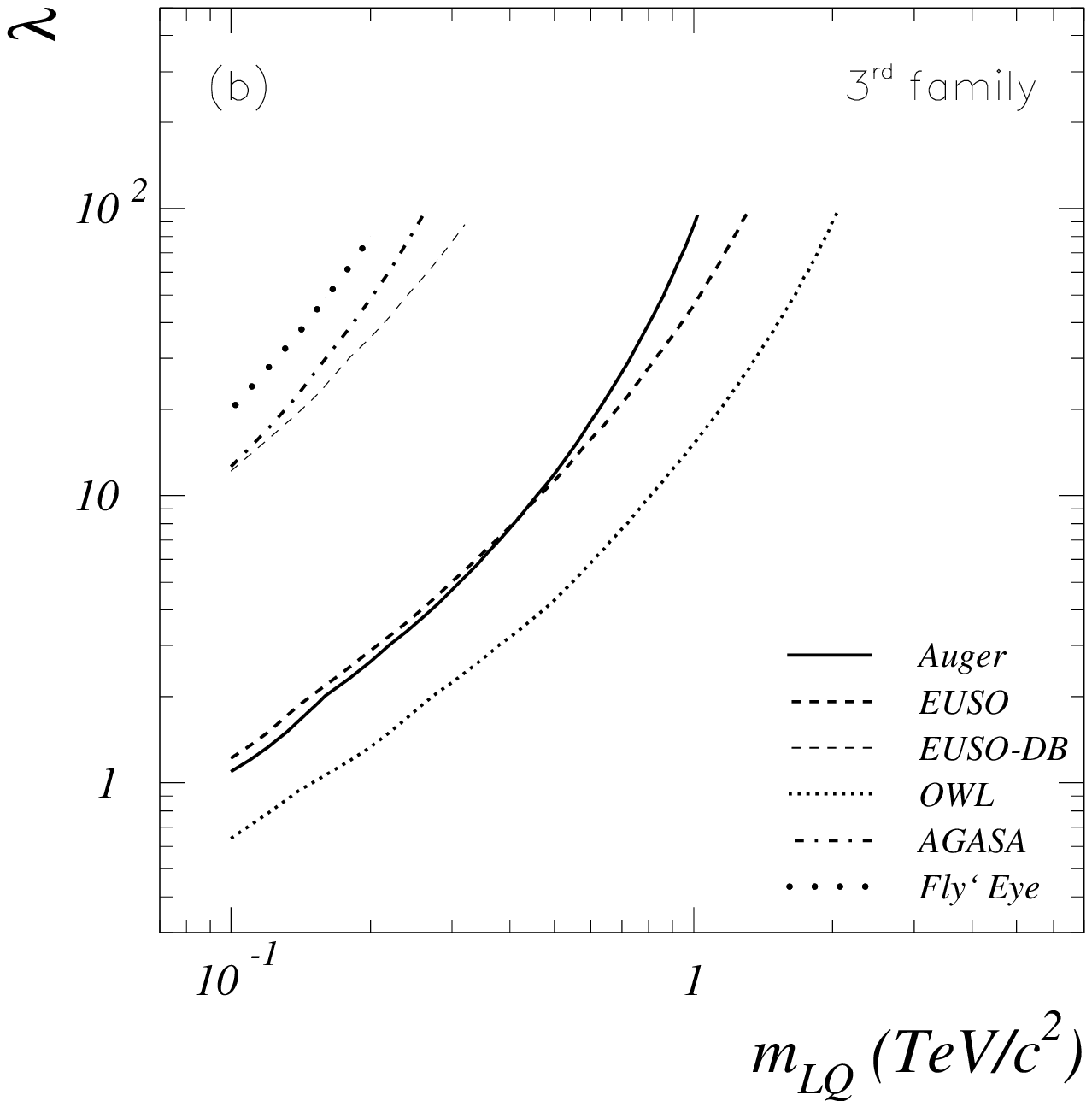}
\end{center}
\caption{Estimated sensitivities of the different cosmic ray
experiments for $S_1$ leptoquarks of second (a) and 
third (b) family 
leptoquarks, as a function of the leptoquark mass.
The observation times were taken as: 10 years for Auger, 
3 years and 10\% duty cycle for both EUSO and OWL. 
For AGASA and Fly's Eye, we followed 
reference~\cite{agasa-a}. In (b), the EUSO-DB line 
shows the expected sensitivity for double bang events (see 
text for details).
}
\label{fig:lims23}
\end{figure}

As discussed above, cosmic ray experiments would allow to search for
leptoquarks of all families, as the initial beam could contain all 
neutrino flavours. We can thus proceed to estimate the expected 
sensitivities for second and third family leptoquarks. These are
shown in Fig.~\ref{fig:lims23}, for scalar leptoquarks, in the 
different considered experiments. In this case, most of the accelerator 
limits discussed above no longer apply, and cosmic ray experiments could
play a role. The D0 limits shown in the figure correspondent to scalar
leptoquarks with both charged and neutral decay mode~\cite{accel}. 
For third generation leptoquarks, the presently available limits  
are typically below 100 GeV/c$^2$.

Furthermore, for third family leptoquarks, an energetic
tau lepton could be produced in the charged decay of a leptoquark.
In these case, the double bang signature proposed in~\cite{dbang,bh-our}
could be searched for: the tau could travel long enough for its
decay to produce a second shower, separate from the first one, 
but still within the field of view of the experiment. 
This rather distinctive new physics signature obviously requires
a very large field of view, while the energy threshold for the
observation of the second bang is also a critical issue. 
In fact, even for the experiments with the largest acceptances,
only a few percent of the detected events are expected to 
have a visible second bang. Using the procedure detailed 
in~\cite{bh-our}, the sensitivity on the leptoquark
coupling as a function of the mass from the observation
of double bang events in EUSO was estimated. This curve 
is also shown in Fig.~\ref{fig:lims23}(b), where we see that
we loose about one order of magnitude with respect to the
sensitivity of the experiment. 

\section{Conclusions}

The sensitivity of very large cosmic ray experiments for the production
of scalar and vector leptoquarks of first, second and third family 
has been explored. While for first family leptoquarks most of the
coupling and mass ranges that could be probed in cosmic ray experiments
have already been excluded by indirect accelerator searches, relevant results
could be obtained for second and third family leptoquarks. Also, 
double bang events could provide a distinctive signature, but only for
very large acceptances or if the fluxes are larger than the ones
considered.



\newpage

\phantom{.}
\vspace{6.cm}


\begin{thebibliography}{00}

\bibitem{agasa}
M. Takeda et al., Phys. Rev. Lett. {\bf 81}, 1163 (1998), 
astro-ph/9807193; \\
N. Hayashida et al., astro-ph/0008102; \\
http://www.akeno.icrr.u-tokyo.ac.jp/AGASA.

\bibitem{fly}
D. J. Bird et al., Phys. Rev. Lett. {\bf 71} (1993) 3401;
Astrophysics J. {\bf 424}, 491 (1994); ibid. {\bf 441}, 144 (1995)
[astro-ph/9410067].

\bibitem{auger}
Auger Collab., {\it The Pierre Auger Project Design Report},
FERMILAB-PUB-96-024, 252 (1996).
http://www.auger.org.

\bibitem{euso}
O. Catalano, In Nuovo Cimento, 24-C, 2, 445 (2001);
http://www.euso-mission.org;
L. Scarsi et al., EUSO Collab., {\it Report on the EUSO Phase A study},
EUSO report EUSO-PI-REP-002-1 (2003).

\bibitem{owl}
J. F. Krizmanic et al., OWL/AirWatch Collab.,
Proc. of the 26$^{th}$ ICRC, Vol. 2, 388 (1999);
http://owl.gsfc.nasa.gov

\bibitem{excit-our}
M.~C.~Espirito Santo, M.~Paulos, M.~Pimenta, J.~C.~Rom\~ao and B.~Tome,
arXiv:hep-ph/0412345.

\bibitem{accel}
C. Adloff et al., .H1 Collab., Phys.Lett. {\bf B568} (2003) 35; 
H1 Collab., arXiv:hep-ex/0506044.
M.Acciarri et al., L3 Collab., Phys. Lett. {\bf B489} (2000) 81;
G. Abbiendi et al., OPAL Collab., Eur. Phys. J. {\bf C6} (1999) 1;
P. Abreu et al., DELPHI Collab., Phys. Lett. {\bf B446} (1999) 62;
V.M. Abazov et al., D0 Collab., Phys. Rev. {\bf D71} (2005) 071104;
C. Grosso-Pilcher et al., for the CDF and D0 Collaborations, hep-ex/9810015;
B. Abbott et al., D0 Collab., Phys. Rev. Lett. {\bf 79} (1997) 4321; 
F. Abe et al., CDF Collab., Phys. Rev. Lett. {\bf 79} (1997) 4327;
D0 Collab., D0 Note 4829.

\bibitem{cosmneut}
F. Halzen and D. Hooper, Rept. Prog. Phys. {\bf 65} (2002) 1025
[astro-ph/0204527];
O.E. Kalashev, V.A. Kuzmin, D.V. Semikoz, G. Sigl, Phys. Rev. 
{\bf D66} (2002) [hep-ph/0205050].

\bibitem{exoticnu}
C. Tyler, A.V. Olinto, G. Sigl Phys. Rev. {\bf D63} (2001)
[hep-ph/0002257];
A. Ringwald, Invited talk at CRIS2004, Catania, May 2004
(2004) [hep-ph/0409151.]

\bibitem{buchmuller:1986zs}
W.~Buchmuller, R.~Ruckl and D.~Wyler,
\newblock Phys. Lett. {\bf B191}, 442 (1987).

\bibitem{agasa-thesis}
L.~A.~Anchordoqui, Z.~Fodor, S.~D.~Katz, A.~Ringwald and H.~Tu,
arXiv:hep-ph/0410136;
H. Tu, PhD thesis, DESY and University Hamburg (2004), DESY-THESIS-2004-018.

\bibitem{wb}
E. Waxman and J. N. Bahcall, Phys. Rev. D {\bf 59}, 023002 (1999)
[hep-ph/9807282]; E. Waxman and J. N. Bahcall, hep-ph/9902383;
K. Mannheim et al., astro-ph/9812398.

\bibitem{agasa-a}
L.~A.~Anchordoqui, J.~L.~Feng, H.~Goldberg and A.~D.~Shapere,
Phys.\ Rev.\ D {\bf 66} (2002) 103002
[arXiv:hep-ph/0207139];

\bibitem{auger-a}
K.~S.~Capelle, J.~W.~Cronin, G.~Parente and E.~Zas,
Astropart.\ Phys.\  {\bf 8} (1998) 321
[arXiv:astro-ph/9801313]; 
L.~A.~Anchordoqui, J.~L.~Feng, H.~Goldberg and A.~D.~Shapere,
Phys.\ Rev.\ D {\bf 65} (2002) 124027
[arXiv:hep-ph/0112247];
L.~A.~Anchordoqui, J.~L.~Feng, H.~Goldberg and A.~D.~Shapere,
Auger Internal Note GAP-2001-053 (2001).

\bibitem{euso-a}
S.~Bottai and S.~Giurgola  [EUSO Collaboration],
Proceedings of the 28th International Cosmic Ray Conferences (ICRC 2003), 
Tsukuba, Japan, 31 Jul - 7 Aug 2003.

\bibitem{owl-a}
S.~I.~Dutta, M.~H.~Reno and I.~Sarcevic,
Phys.\ Rev.\ D {\bf 66} (2002) 033002
[arXiv:hep-ph/0204218].

\bibitem{dbang}
D. Fargion et al., astro-ph/9704205 (1997);
D. Fargion et al., astro-ph/0305128 (2003);
S. Bottai and S. Giurgula, astro-ph/0205325 (2002);
H. Athar, G. Parente, E. Zas, Phys.Rev. {\bf D62}, 093010 (2000), hep-ph/0006123; 
M. M. Guzzo and C. A. Moura Jr., hep-ph/0312119.

\bibitem{bh-our}
V.~Cardoso, M.~C.~Espirito Santo, M.~Paulos, M.~Pimenta and B.~Tome,
Astropart.Phys. {\bf 22} (2005) 399.

\end{thebibliography}
\end{document}